\documentclass[11pt,a4paper]{article}
\usepackage[latin1] {inputenc}
\usepackage{amsmath}
\usepackage{amsfonts}
\usepackage{amssymb}
\usepackage{enumerate}

\usepackage{tikz}
\usetikzlibrary{arrows,decorations.pathmorphing,backgrounds,fit}
\usepackage{pgfmath}
\usepackage{sidecap}

\newtheorem{definition}{Definition}

\newtheorem{lemma}[definition]{Lemma}
\newtheorem{theorem}[definition]{Theorem}
 
\newtheorem{cor}[definition]{Corollary}
\newtheorem{obs}[definition]{Observation}

\newtheorem{claim}[definition]{Claim}

\newcommand\dw[2]{\draw[#1!#2,fill=#1!#2]}

\newcommand{\Prob}[1]{\mathbf{P} \left( #1 \right)}

\newcommand{\Probc}[2]{\mathbf{P} \left( #1 \;\left|\; #2 \right.\right)}

\newcommand{\sCZ}{\mbox{\sc{CZ}}}
\newcommand{\sa}{\textsc{a}}
\newcommand{\sbb}{\textsc{b}}
\newcommand{\dx}{\mbox{d$x$}}
\newcommand{\dy}{\mbox{d$y$}}

\newcommand{\sP}{\sc{P}}
\newcommand{\sv}{\mbox{\sc{v}}}

\newcommand{\proof}{ \noindent \textit{Proof. }}
\newcommand{\qed}{\hspace{\stretch{1}$\square$}}

\title{\textbf{ Fast Flooding over Manhattan }}

\author{ 
Andrea Clementi\thanks{Contact Author:  \emph{Tor Vergata} University of Rome  (clementi@mat.uniroma2.it).} 
 \and Angelo Monti\thanks{\emph{La Sapienza} University of Rome   (\{monti,silver\})@di.uniroma1.it.  } 
  \and Riccardo Silvestri$^{\dagger}$ }

\begin{document}
\maketitle

\begin{abstract}
We consider a \emph{Mobile Ad-hoc NETwork (MANET)} formed by  $n$ agents that move
at speed $\sv$ according to the \emph{Manhattan Random-Way Point} model  over a
square region of side length $L$. The resulting  stationary (agent) spatial probability distribution is far 
to be uniform: the average density over the ``central zone" is asymptotically higher than that over 
the ``suburb". Agents exchange data iff they are at distance at
most $R$ within each other. 

 We study the \emph{flooding time} of this MANET: the number of time steps
required to broadcast a message from one source agent to all agents of the network in the stationary phase.
We prove the first asymptotical upper bound on the flooding time.  This bound  holds with high 
probability, it is a decreasing function of $R$ and $\sv$, and it
is tight for a wide and relevant range of the network parameters (i.e. $L$, $R$ and $\sv$).

A consequence of our result is that  flooding  
over the  sparse and  highly-disconnected suburb   
can be as fast as flooding over the dense and connected central zone.
Rather surprisingly, this property holds even when $R$ is exponentially
 below the connectivity threshold of the MANET and the speed $\sv$ is very low.

\end{abstract}

\bigskip
\noindent
\textbf{Keywords:}    Mobile Ad-Hoc Networks, Flooding Protocols, Probabilistic Analysis.

\thispagestyle{empty}

\newpage

\section{Introduction}

We continue our adventure in exploring the impact of agent  mobility
on data propagation in \emph{Mobile Ad-hoc NETworks} (\emph{MANET}).
Node mobility can be considered as a resource to exploit in data-forwarding protocols rather than a
hurdle \cite{KN08,JetAl06}. This is well-captured by the model known as  
opportunistic MANET \cite{PPC06,GT02,JetAl06,ZAZ04}.  
 
In our previous works \cite{CMPS09} and  
\cite{CPS09}, the speed of information spreading (measured in terms of flooding time) has
been analytically determined in MANETs where agents perform a sort of  independent
random walks over a square. In  such MANETs,  the stationary (agent) spatial  probability  distribution is
almost uniform:  the probability that an agent lies in a given position
is almost the same for any choice of the position. Furthermore, the
stationary (agent) destination probability distribution (i.e., the probability that an agent, lying in a
given position $(x,y)$, has a given destination) is uniform over a disk centered on $(x,y)$
and it is zero elsewhere.   

The most popular mobility model is the \emph{Random Way-Point} (RWP) \cite{BRS03,CBD02,LV06}.
In the basic version of this model, each agent chooses independently and uniformly at random a 
destination over all the square. Then, she starts traveling at speed $\sv$ towards the destination 
along a simple path.  When she reaches the destination (a way-point),  she chooses another 
destination.

In this work, we   consider the version of the RWP, called \emph{Manhattan
Random Way-Point} (MRWP) model,  where the path followed by an agent from a given point 
to a destination point is randomly chosen between the two Manhattan shortest paths connecting 
the two points (see Section~\ref{sec::manh} for a formal definition). 
This version of the RWP is motivated by scenarios where agents travel over an urban zone 
and try to minimize the number of \emph{turns} while keeping the chosen route as short as 
possible \cite{CDMRV09,CMS10a,LV06}.  

The MRWP is a dynamical system yielded by an infinite Markovian  process that always reaches 
a stationary phase. Explicit formulas for the stationary spatial and destination probability distributions 
have been derived in \cite{CDMRV09,CMS10a}. The stationary distributions of some other versions of 
the RWP have been obtained in \cite{BRS03, CNB04,LV05,L07}. In general,
the knowledge of such distributions is crucial to achieve perfect simulation \cite{CBD02,LV06}, 
to derive connectivity properties \cite{DMP08,CDMRV09}, and for the study of information spreading 
\cite{CMPS09,CPS09,CDMRV09}.

Differently from the case of random-walks models,  it turns out that the stationary probability distributions
yielded by the MRWP are very far to be uniform.
As for the stationary  spatial distribution, in the four regions close to the corners of the square, the
probability  density function is asymptotically much lower than that in the \emph{Central Zone} 
(see Fig.~\ref{fig_manhattan}). The four corners form the \emph{Suburb}. The area of the
Suburb is not negligible, being a constant fraction of the entire area. A
further crucial difference with respect to the random-walk models, studied in
\cite{DMP08,CMPS09,CPS09}, lies in the stationary destination distribution. 
This distribution has a rather complex shape (see Section~\ref{sec::manh}).   


We consider a MANET where $n$ agents move independently at speed $\sv >0$ over a square 
of side length $L$ according to the MRWP model. Agents exchange data iff they are at distance at
most $R$ within each other, where $R>0$ is the \emph{agent transmission radius}. 
At every time step $t$, the snapshot of the MANET determines
a symmetric disk graph $G_t$.  The \emph{connectivity threshold} is the smallest $R$ such that 
$G_t$ is connected.  In \cite{CDMRV09}, from the explicit formula of stationary spatial distribution 
of the MRWP model, the relative connectivity threshold for the stationary graph  $G_t$ is
derived:  when $L = \sqrt n$, it is equal to some root of $n$. Thus, it is exponentially higher than that 
of the stationary disk graph yielded by mobility models having uniform stationary spatial 
distribution (such as the random-walk models and some RWP models over toroidal spaces):
this threshold being $\Theta(\sqrt{\log n})$ \cite{GK99,Penrose}.
 
The \emph{flooding mechanism} is the simple broadcast protocol
where every informed agent sends the source message at discrete time steps 
(an agent  is said to be informed if she knows the source message). 
Thus, a non-informed agent $\sa$ gets informed at time step $t$ iff, during $t$, an 
informed agent $\sbb$ is within distance $R$ from $\sa$.
The \emph{flooding time} is the first time step in which all agents are informed.
It is a natural lower bound for any broadcast protocol, it represents the maximal
speed of data propagation, and it has the same role of the diameter in static networks. 
Flooding time of some classes of Markovian evolving graphs \cite{AKL08} has been recently 
studied in \cite{CMMPS08,CMPS09,BCF09}. On the other hand, no analytical results are known 
for flooding time on any version of the RWP model.

\smallskip
\noindent \textbf{Our  Result.} 
We study the flooding time in the MRWP model under the conditions: $R \geqslant c_1 \sqrt {\log n} $  
and $\sv \leqslant \ R / c_2 $, where $c_1$  and $c_2$ are positive constants. Observe that, from the above 
discussion on graph connectivity, the first assumption on $R$  does not guarantee network
connectivity \cite{CDMRV09}:  the stationary snapshots could  be   highly disconnected in the Suburb.
The second assumption means we are considering a slow-mobility scenario, i.e.,
 when an agent's move, in a time unit, cannot be longer than
 the  transmission radius. Observe that the second assumption implies the lower bound 
 $\Omega(L/R)$ for flooding time.
 We prove that flooding time is w.h.p.\footnote{As usual, 
we say event $\mathcal E$ holds \emph{with high probability (w.h.p.)} if $\Prob{\mathcal E} \geqslant 1 - 1/ n^{c}$ for 
some $c >0$.} bounded by
 \begin{equation}\label{eq::intromain}
 {\large O}\left(\frac{L}{R} + \frac{S}{\sv}\right)
 \end{equation}
 where 
 \[
 S = \Theta\left(\frac{L^3 \log n}{R^2 n}\right)
 \]
 is the diameter of each of the four corner regions of the Suburb. Hence, our bound can be interpreted
 as saying that flooding time is asymptotically bounded by the sum of two consecutive 
 time spans: the time to
 traverse the square at  ``speed'' $R$ and the   time to traverse   the Suburb at speed $\sv$.

 Let us discuss the consequences of our  bound in the
 ``standard'' case $L = \sqrt n$  (clearly, similar consequences hold for 
 different values of $L$).  If  $R= \Theta(\sqrt{\log n})$, our bound becomes
 $O(L/\sv)$:  this is optimal whenever  $\sv = \Theta(R)$. In general,
 our bound is optimal whenever the speed $\sv$ falls into the 
 range $ \frac{\log n} R   \leqslant \sv \leqslant R$. 
 For instance, if $R = \Theta(\log n)$, our bound becomes optimal
 provided that $\sv$ is larger than an absolute constant.  This fact  is rather
 surprisingly. Indeed, under such  conditions, the snapshots in the Suburb are sparse 
 and highly-disconnected (as mentioned above, the connectivity threshold in the Suburb
 is exponentially larger than $\log n$  \cite{CDMRV09}). Nevertheless, our
 bound says   that  flooding  succeeds over  the Suburb   as well and,   even more,
 its completion time is similar to that in the Central Zone where the snapshots are fully-connected.
This phenomenon holds even when the agent speed $\sv$ is very low.

\noindent
We do not  know whether our bound is optimal for all the range of the network parameters, however it cannot
be improved to $O(L/R)$.
Indeed, we prove that, for some ranges of $R$, $n$, and $L$, flooding time is  asymptotically
larger  than $L/R$ and, moreover, it must depend on $\sv$.

Finally,  
we strongly believe that our  ideas and techniques used to obtain 
 the  upper bound can be adapted to analyze   flooding over 
other versions of the RWP model and even over some versions 
of the more general Random Trip model \cite{LV06}, in the stationary phase. 
An outline of  our proof and its potential applications to other mobility models are  given in
Section \ref{sec::overview}.

\smallskip
\noindent
\textbf{Note.} In order to make the technical arguments more readable, our asymptotic analysis definitely  
does not optimize the constants in the upper bound and in the assumption on $R$. 
However, we believe that, with some work, our arguments could be refined so that   
  the involved constants are significantly improved.

  \section{The Manhattan Random-Way Point} \label{sec::manh}


In this section, we formally  present  the  MRWP model.
Consider a square of edge length $L>0$. A set  of $n$ independent \emph{agents}
move over this square according to the following stochastic rule.
Starting from an initial   position $(x_0,y_0)$,  every agent  selects a   \emph{destination}  $(x,y)$ uniformly at random
in  the square. Then, the
agent chooses uniformly at random  between  the two    feasible \emph{Manhattan paths}
\[  \sP_1 = ((x_0,y_0) \rightarrow (x_0,y) \rightarrow (x,y))   \  \mbox{ and } \
 \sP_2 = ( (x_0,y_0) \rightarrow  (x,y_0) \rightarrow   (x,y)  )  \]
 \noindent
 Once  the feasible path is selected,   the agent starts  following  the chosen route
 with  \emph{constant   velocity} determined by the parameter $\sv$.
  We assume that all agents have the same velocity $\sv$ that represents
     the travelled distance by an agent in the time unit.
     An agent, once arrived at the selected destination, re-applies  the   process described above again and again. This  Markovian process yields the MRWP model.
     
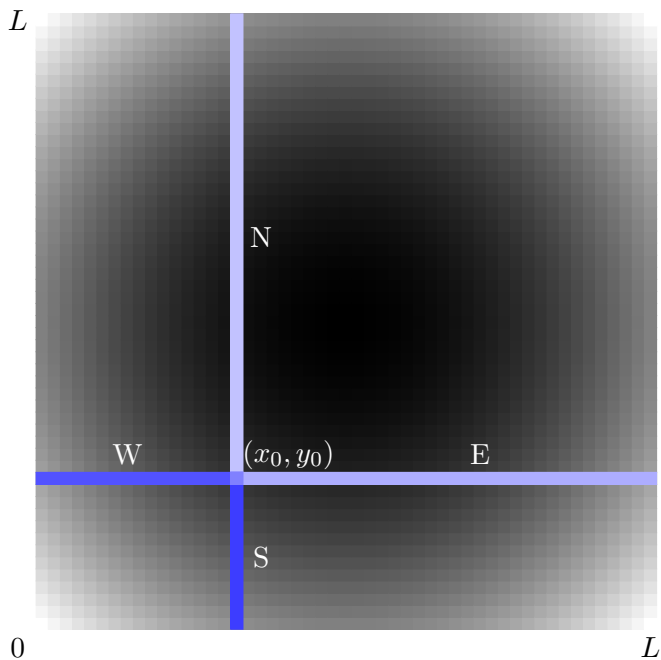
\begin{SCfigure}
\begin{tikzpicture}
[scale=0.8]
\def\LL{10}
\pgfmathsetmacro{\LLQ}{200/(\LL*\LL)}
\def\list{0,.2,...,\LL}
\foreach \x in \list {
    \foreach \y in \list {
        \pgfmathsetmacro{\d}{(\LLQ*(\LL*(\x + \y) - \x*\x - \y*\y))}
        \dw{black}{\d} (\x, \y) +(-.1, -.1) rectangle ++(.1, .1);
    }
}
\def\dest{blue}
\pgfmathsetmacro{\NN}{100/\LL}
\pgfmathsetmacro{\XX}{floor((\LL/0.2)/3)*0.2}
\pgfmathsetmacro{\YY}{floor((\LL/0.2)/4)*0.2}
\pgfmathsetmacro{\d}{\NN*\YY}
\pgfmathsetmacro{\LY}{\LL + .1 - \YY}
\dw{\dest}{\d} (\XX, \YY) +(-.1, -.1) rectangle ++(.1, \LY); 
\pgfmathsetmacro{\d}{\NN*(\LL - \YY)}
\pgfmathsetmacro{\YYZ}{\YY + .1}
\dw{\dest}{\d} (\XX, \YY) +(-.1, -\YYZ) rectangle ++(.1, .1); 
\pgfmathsetmacro{\d}{\NN*(\LL - \XX)}
\pgfmathsetmacro{\XXZ}{\XX + .1}
\dw{\dest}{\d} (\XX, \YY) +(-\XXZ, -.1) rectangle ++(.1, .1); 
\pgfmathsetmacro{\d}{\NN*\XX}
\pgfmathsetmacro{\LX}{\LL + .1 - \XX}
\dw{\dest}{\d} (\XX, \YY) +(-.1, -.1) rectangle ++(\LX, .1); 
\dw{\dest}{50} (\XX, \YY) +(-.1, -.1) rectangle ++(.1, .1); 
\pgfmathsetmacro{\XXW}{\XX + .85}
\pgfmathsetmacro{\YYW}{\YY + .4}
\draw[white] (\XXW, \YYW) node {$(x_0, y_0)$};
\pgfmathsetmacro{\XXW}{\XX + 4}
\draw[white] (\XXW, \YYW) node {E};
\pgfmathsetmacro{\XXW}{1.4}
\draw[white] (\XXW, \YYW) node {W};
\pgfmathsetmacro{\XXW}{\XX + .4}
\pgfmathsetmacro{\YYW}{\YY + 4}
\draw[white] (\XXW, \YYW) node {N};
\pgfmathsetmacro{\YYW}{\YY - 1.3}
\draw[white] (\XXW, \YYW) node {S};
\draw (-.4, -.4) node {$0$};
\draw (\LL, -.4) node {$L$};
\draw (-.4, \LL) node {$L$};
\end{tikzpicture}
\caption{The spatial  density function   is shown by a gradation 
of gray (black corresponds to the maximum density and white corresponds to the 
minimum density).  The destination probability over the cross 
of   agent   position $(x_0, y_0) = (L/3, L/4)$ is shown in gradation of blue.}   \label{fig_manhattan}
\end{SCfigure}

     \noindent
       The stationary   probability distributions of 
     the MRWP have been recently analytically derived.
           The   \emph{stationary
     (agent) spatial   distribution} gives the probability that an agent lies in a position $(x,y)$ and  it has been derived in 
     \cite{CDMRV09}.
     The  \emph{ stationary (agent) destination distribution}   gives the probability that an agent, conditioned to lie in position
     $(x_0,y_0)$,   is traveling  toward     destination $(x,y)$ and it  has been  determined in 
     \cite{CMS10a}.  An informal representation of the two distributions is given in Fig. \ref{fig_manhattan}.

\noindent

     \begin{theorem}\label{thm::manhspat}  \cite{CDMRV09}.
       The     probability density function of the  stationary  spatial distribution  is    
\begin{equation}\label{rpd}
f(x,y) \ = \  \frac{3}{L^3}(x+y)-\frac{3}{L^4}\left({x}^2+{y}^2\right)  
\end{equation}
\end{theorem}
      
\begin{theorem} \label{thm::manhdest1} \cite{CMS10a}.
The probability density function of the  stationary  destination distribution is
 
{\small \begin{equation} \label{eq::DENSITY}
 f_{(x_0,y_0)}(x,y) =\left\{
\begin{array}{ll}
  \frac {2L - x_0 - y_0}{4L(L(x_0+y_0) -(x_0^2 +y_0^2 ))} & \mbox{if $x<x_0$ and $y<y_0$}\\
\frac { x_0+y_0} {4L(L(x_0+y_0) -(x_0^2 +y_0^2 ))}     & \mbox{if $x>x_0$ and $y>y_0$}\\
     \frac { L-x_0 + y_0 }{4L(L(x_0+y_0) -(x_0^2 +y_0^2 ))} & \mbox{if $x<x_0$ and $y>y_0$}\\
  \frac { L + x_0 -y_0}{4L(L(x_0+y_0) -(x_0^2 +y_0^2 ))} & \mbox{if $x>x_0$ and $y<y_0$}\\
  + \infty& \mbox{otherwise.}
\end{array}
\right.
\end{equation}
}\end{theorem}

\noindent
In order  to get  the   destination distribution where the probability density function is infinite,   one has to consider the four segments outgoing from
$(x_0,y_0)$ and parallel to the axis (see Fig. \ref{fig_manhattan}). 
For every \emph{segment} $\sc{s} \in  \{ \mbox{S}, \mbox{W}, \mbox{N}, \mbox{E} \}$,  
  the probability that an agent in  node $(x_0,y_0)$, 
  has destination lying on  the segment $\sc{s}$, is $\phi^{s}_{(x_0,y_0)}$. It  
 has been proven in \cite{CMS10a} that 

\begin{equation}\label{eq::fisouth}
 \phi^{\mbox{S}}_{(x_0,y_0)} \ =  \ \phi^{\mbox{N}}_{(x_0,y_0)} \  = \
\frac{y_0(L-y_0)}{4L(x_0+y_0) - 4(x_0^2 + y_0^2)}
 \end{equation}

\begin{equation}\label{eq::fiwest} \phi^{\mbox{W}}_{(x_0,y_0)} \ =  \ \phi^{\mbox{E}}_{(x_0,y_0)} \   = \
\frac{x_0(L-x_0)}{4L(x_0+y_0) - 4(x_0^2 + y_0^2)}
 \end{equation}

\noindent
It is interesting to observe that  the sum of the above four probabilities (i.e. the probability that the agent has destination
over the \emph{cross} centered on $(x_0,y_0)$) is not zero (it is equal to $1/2$) despite the fact that this region (i.e. the cross) has area 0.
This fact will be used  in our  analysis of   flooding over the Suburb.

 \section{Proof of the upper bound: an overview}\label{sec::overview}
 
  We here outline the proof technique of the upper bound on the 
flooding time. The stationary spatial probability distribution (see Fig. \ref{fig_manhattan}) 
shows a central region of high density (the Central Zone) and 
four corner regions (the Suburb) of low density. High density 
means that the   expected number of agents in any disk of radius $R$ 
is  $\Omega(R^2)$  (see Def.~\ref{def::centralzone}) and that, in the Central Zone, 
the resulting MANET is w.h.p. connected.
  
A key-issue here is
 that flooding  must be observed over a sequence of consecutive snapshots of the MANET:  even though we can
say that each of such snapshots   (individually) enjoys    all the stationary  properties (such as high density and connectivity of Central Zone),  
we cannot directly exploit them during the observed process.
Indeed, there is  strong stochastic dependence between  consecutive snapshots: if we ``observe" one snapshot, then the next one is not anymore
random with   the stationary distribution. 

\noindent
Informally  speaking, this technical issue is solved by proving  that  \emph{any} stationary snapshot
sequence of reasonable length (say $O(n)$) is formed by  \emph{conditional}   
random graphs, all having   expansion properties similar to those
enjoyed by  the (individual) stationary random graph.   Then,  
the Central Zone is partitioned 
in square cells of side length less than $R$ and the flooding process on agents of the MANET
is viewed as a propagation of information from cells to their 
adjacent ones (see Lemmas \ref{stability}, \ref{boundary}, and Theorem \ref{spreading}).
  Thanks to the above expansion properties, we prove   this 
propagation takes $O(L/R)$ time to ``infect'' all the Central Zone. 
 We  introduced  a  similar technique  in \cite{CMPS09} for the random-walk mobility model:
  we here carefully adapt it for the particular
shape of the Central Zone and for the different random agent mobility (i.e. the  MRWP model).

The analysis of the flooding over the Suburb is much harder. Indeed, 
besides the above  key-issue,  in the sparse and
highly disconnected Suburb,
we   cannot
exploit any good expansion property of the snapshots.    Moreover,  it is possible to prove that, with non negligible probability,
there are  agents that will not visit the Central Zone for a (too) long  time.   
However, we resort to the properties of the stationary  destination probability 
distribution (combined to  those of the spatial distribution) 
to prove that a sufficiently wide flow of informed agents  floods from the 
Central Zone over the Suburb. This is not enough 
to guarantee that all the agents in the Suburb gets informed within 
$O(S/\sv)$ time after the Central Zone is informed. A special care 
has indeed  to be used to guarantee that this  flow of informed agents 
floods over the Suburb \emph{sufficiently fast} (see Lemma \ref{lm:meeting time}).

We believe that the above outlined technique can be adapted 
to analyze the flooding over other versions of the RWP model 
and    over  some versions of the more general Random Trip 
model \cite{LV06}. Indeed, in several versions of those models the spatial 
probability distribution shows a high-density region and low-density 
regions. It should not be too hard to adapt our analysis 
  for  the high-density region. 
The analysis for  the low-density regions remains the hard part. 
However, we think that, by following our proof (Lemmas \ref{lem:suburb1}, \ref{lem:suburb2}, and \ref{lm:meeting time}) as a 
guideline and making the most of the properties of 
both distributions (destination and spatial), the flooding 
over the Suburb can be analyzed as well.

\section{Flooding   over   Manhattan} \label{sec::flooding}

 We  consider  the   \emph{flooding} process
 starting from a source agent. We want to provide
an   upper  bound on the  \emph{flooding time}, i.e. the time
 required by the flooding process to inform all the      agents.
  An agent is said to be \emph{informed}    if she knows 
   the source message.
 At the starting time, only the source agent is informed. Then, an agent $\sa$
 gets informed at time step  $t = 1,2, \ldots $ iff, at that time,
 there is (at least)  an informed agent $\sbb$ that lies at distance not larger than
 $R >0$, where $R$ is  the transmission radius valid for all agents.
 \noindent
 The $n$ agents move over the    square $L \times L$  (with $L >0$)  according to the MRWP model.
 This section is devoted to prove our main  result.
  
\begin{theorem}\label{thm:main}
Consider a MANET of   $n$ agents moving over a square of size $L$ according to
the MRWP model. Let $R$ be the transmission radius and $\sv$ be the agent speed.
Assume that $R \geqslant c_1 L \sqrt{\log n / n}$ and $\sv \leqslant R/c_2 $ for sufficiently large constants
$c_1$ and $c_2$. Then,  for sufficiently large $n$, the flooding time in the stationary phase 
 is w.h.p. bounded by

\[    {\large O} \left(   \frac LR +  \frac{L}{\sv} \frac{L^2}{R^2} \frac{\log n}{n}     \right)  \]
\end{theorem}

  The proof of the above theorem requires some preliminary definitions and results.
  Notice that, \noindent
due to lack of space,  most of   proofs of the next  lemmas are given in the Appendix.

  \noindent
    In the sequel,  we  assume that     $R  \leqslant   \sqrt 2 \, L$. Observe that  if $R >\sqrt 2 \, L$, then the bound  on the flooding time is trivial. 
We partition the  square  into  $m \times m$ square   \emph{cells} of  side length $\ell$
   with
\begin{equation}\label{primo1}
\frac{R}{1 + \sqrt 5} \leqslant   \ell \leqslant \frac{R}{\sqrt{5}}
\end{equation}
Notice that    $\ell$ is chosen in order to guarantee that an agent inside a cell $C$ can transmit   to any 
agent lying in any  of the four adjacent cells of $C$.  The \emph{core} of a cell $C$ is the central subsquare of $C$ with side length $\ell/3$.
We    assume   the agent transmission radius and the agent speed   satisfy  the following bounds

 \begin{equation} \label{eq::TransmBound}
  R \geq 200 L \sqrt{\frac{\log n}{n}}  
 \end{equation}

\begin{equation} \label{eq::SpeedBound}
 \sv   \le  \frac R{ 3(1 + \sqrt 5) }    \end{equation}

\noindent
Observe that the  above condition on $\sv$  guarantees    that an agent, lying in the core of a cell $C$ at time $t$, 
 will remain  in $C$ at time $t+1$ as well.

\noindent
From Eq. \ref{rpd}, the   probability that an agent lies in a cell $C$ is given by

\[ \int_C  f(x,y) dx   dy  \]
We now define formally the   \emph{Central Zone} and the  \emph{Suburb} 

\begin{definition} \label{def::centralzone} \textbf{[Central Zone and Suburb]}
The
   \emph{Central Zone}  is the
 subset $\sCZ$ of those cells   $C$ such that 
 \[ \int_C  f(x,y) dx   dy \  \geqslant \ \frac{3}{8}\,  \frac {\log n} n \] The complement set of the Central Zone is called
 \emph{Suburb}.
   \end{definition}


\medskip
\noindent
{\large {\bf Flooding in the Central Zone}}

   \noindent By using the density function $f(x,y)$ in  Eq. \ref{rpd} and Ineq. \ref{primo1},  we have 
  \begin{obs}  \label{obs::cellprob}
Consider a cell $C$  having  its   South-West (SW) corner  in position $(x_0,y_0)$.
Then, it holds that
\footnotesize
\[    \int_C  f(x,y) dx   dy  =  \frac{3\ell^2}{L^4} \left(   \frac {\ell} 3  (3L - 2\ell) 
+ x_0(L-\ell - x_0) +  y_0(L -\ell -y_0) \right)  \  \geqslant \  \frac{\ell^3 (3L -2 \ell)}{L^4}\geqslant \left(\frac{R}{(1+\sqrt 5) L}\right)^3 \]
\normalsize
\end{obs}
    
    \noindent  From the above observation, it is easy  to verify that the constant 
   3/8  in Def. \ref{def::centralzone}   guarantees the following

     \begin{lemma} \label{obs::38}
   The  number  
  of   rows (and   columns as well) of cells that   belong to the Central Zone is at least $m /\sqrt 2$.
  \end{lemma}

\noindent We say that the \emph{density condition} holds at time $t$ if, for every cell $C$ of the Central Zone,
the number of agents in the core of  $C$ at time $t$ is at least $\eta \log n$, for a suitable positive constant.
 Let $\mathcal{D}$ be the following event:  the density condition
  holds for every time step $t = 0, 1,\ldots, n$.

\noindent The proof of the following lemma easily derive from  the definition of  the Central Zone by a simple union bound argument.

\begin{lemma}[Density] \label{lm::density}
 The probability  of  event $\mathcal{D}$ is at least $1-1/n^4$.
\end{lemma}
 
\noindent In the analysis of the flooding over the Central Zone,
 we will tacitly  assume that  event $\mathcal{D}$ holds.
Thanks to the previous lemma, since we are conditioning w.r.t.  an event that holds w.h.p., the
corresponding unconditional     probabilities are affected by a negligible factor only.

\noindent
We say that a cell $C$ is \emph{informed} at time $t$ if all   agents visiting  $C$ at time $t$ are informed.
Consider an informed cell $C$ of the Central Zone at time $t$. By the density condition, its core contains at least one informed agent  $\sa$. Thanks to the Ineq. \ref{eq::SpeedBound}, agent $\sa$  will remain inside
$C$ during all time step $t+1$. Then, after the agent transmission   of time $t+1$, all agents   lying in $C$ or  in its   adjacent cells at time $t+1$,  will get informed.
We have thus shown the following

  \begin{lemma} [Stability] \label{stability}
  For any $0 \le t \le n$, If a cell $C$ of the Central Zone is informed at time $t>0$, then  $C$ and all its adjacent cells
  (in the Central Zone)  will be  informed at time $t +1$
  with probability at least $1 - 1/n^4$.
  \end{lemma}

 \noindent
  For any subset $B$ of   cells of the Central Zone, define the \emph{boundary}  $\partial B$ of $B$ as follows
  \[ \partial B = \{ C \;|\; C \in \sCZ \setminus B \;\wedge\; \exists C'\in B:  \mbox{ $C'$ is adjacent to $C$}\}. \]

\noindent
We now provide a lower bound on the expansion of \emph{any} cell subset of the Central Zone.
The result is   similar to that in 
 \cite{CPS09},  however, in our case, the proof  must be adapted for the particular   shape of the Central Zone.


\begin{lemma}[Boundary]\label{boundary}
Let $B$ be any cell subset of the Central Zone.  It holds that
\[ |\partial B| \;\geqslant\; \sqrt{\min\{|B|, |\sCZ| - |B|\}}. \]
\end{lemma}
 
 \noindent
By exploiting Lemmas~\ref{stability}, \ref{boundary}  we get the following bound.

\begin{theorem} \label{spreading}
Assume that, at time $t=0$,    at least one informed agent lies in the Central Zone.
 Then, with probability at least $1 -  1/n^2$,
at every time $t$ with $18\frac{L}{R} \leqslant t \leqslant n$, all the cells in the Central Zone
are informed.
\end{theorem}

\proof For any $t \geqslant 0$, let $\mathcal{Q}_t$ be the set of  informed  cells  at time $t$ in the Central Zone.
By hypothesis
$|\mathcal{Q}_0| \geqslant 1$. In virtue of Lemma~\ref{boundary}, if all cells in $\mathcal{Q}_t$
and all their adjacent cells get  informed at time $t+1$, then
$ |\mathcal{Q}_{t+1}| \geqslant |\mathcal{Q}_t| + \sqrt{\min\{ |\mathcal{Q}_t|, |\sCZ| -   |\mathcal{Q}_t| \} }$.\\
 This implies that if the above inequality does not hold then
    a cell
$C\in\mathcal{Q}_t$ exists  such that   $C$ or one adjacent cell of $C$ is not  informed at time $t+1$.
It follows that
\[ \Prob{|\mathcal{Q}_{t+1}| < |\mathcal{Q}_t| + \sqrt{\min\{ |\mathcal{Q}_t|,  |\sCZ| -  |\mathcal{Q}_t| \} }}  \;\leqslant\;
\Prob{\exists C\in \mathcal{Q}_t : \mathcal{E}_{C, t+1}} \]
where $\mathcal{E}_{C, t+1}$ is the event that occurs if $C$ or one of its adjacent cells in $\sCZ$  is not  informed at time $t+1$. By the union bound, it holds that
\begin{small}
\[
\Prob{\exists C\in \mathcal{Q}_t : \mathcal{E}_{C, t+1}} \;\leqslant\; \sum_{C} \Prob{C\in \mathcal{Q}_t
\wedge \mathcal{E}_{C, t+1}} \;=\; \sum_{C} \Probc{ \mathcal{E}_{C, t+1} }{ C\in \mathcal{Q}_t }
\Prob{C\in \mathcal{Q}_t}.
\]
\end{small}
From Lemma~\ref{stability}, for every cell $C$, we have 
$\Probc{\mathcal{E}_{C, t+1} }{C\in \mathcal{Q}_t } \leqslant 1/{n^4}$.
It follows that
\begin{eqnarray*}
\Prob{|\mathcal{Q}_{t+1}| < |\mathcal{Q}_t| + \sqrt{\min\{ |\mathcal{Q}_t|, |\sCZ| -  |\mathcal{Q}_t| \} }}
 & \leqslant & \sum_{C} \Probc{\mathcal{E}_{C, t+1} }{C\in \mathcal{Q}_t }
\Prob{C\in \mathcal{Q}_t} \\
 \leqslant   \sum_{C} \Probc{\mathcal{E}_{C, t+1} }{ C\in \mathcal{Q}_t }
& \leqslant & \frac{|\sCZ|}{n^4} \;\leqslant\; \frac{1}{n^3}.
\end{eqnarray*}
Thus, by the union bound, with probability at least $1 - \frac{1}{n^2}$, it holds that
\[ \forall t = 0,1,\ldots, n \qquad |\mathcal{Q}_{t+1}| \;\geqslant\; |\mathcal{Q}_t| + \sqrt{\min\{ |\mathcal{Q}_t|, |\sCZ| -  |\mathcal{Q}_t| \} }. \]
\noindent Now we use the following claim proven in \cite{CPS09}

{\small  \begin{quote}
\begin{claim}
Let $\bar q$ be any integer s.t. $\bar q \geqslant 1$ and let $\{q_t \;|\; t\in \mathbb{N}\}$ be a sequence of integers such that
$q_0 \geqslant 1$, for every $t\geqslant 0$, $q_t \leqslant \bar q$ and $q_{t+1} \geqslant q_t + \sqrt{\min\{q_t, \bar q - q_t\}}$.
Then, it holds that, for every $t \geqslant 5\sqrt{\bar q}$, $q_t = \bar q$.
\end{claim}
\end{quote}
}
\noindent
By applying the claim  with $q_t = |\mathcal{Q}_t |$ and $\bar q = |\sCZ|$, we get
$|\mathcal{Q}_t| = |\sCZ|$    for every $t$ with
$5\sqrt{|\sCZ|} \leqslant t \leqslant n $. The thesis follows 
   since  $5 \sqrt{|\sCZ|} \leqslant  5 L/\ell \leqslant 18 L/R$.

\qed

\noindent
Notice  that, thanks to Obs. \ref{obs::cellprob},  if $R  \geqslant \frac{(1 + \sqrt 5)}{2}  L \left(\frac{3\log n}{ n}\right)^{1/3}$ , then       \emph{all} cells of the support square belong to the Central Zone.  
Hence, from Theorem \ref{spreading}, we get

\begin{cor} [Large  $R$]  \label{cor::allcentr}
If $R \geqslant  \frac{(1 + \sqrt 5)}{2}  L \left(\frac{3\log n}{ n}\right)^{1/3}$, with probability at least $1 - 1/n^2$,
the   (overall)  flooding time is   $18 L / R$.
\end{cor}

\medskip


\noindent
 {\large {\bf Flooding over the Suburb}}
 
 \noindent
 We now  analyze the flooding process over   the  Suburb.
 Thanks to Cor. \ref{cor::allcentr},   we can assume that 
 
 \begin{equation} \label{eq::sub1}
 R \leqslant \frac{1 + \sqrt 5}{2}  L \left(\frac{3\log n }{ n}\right)^{1/3}
 \end{equation}
 otherwise the Suburb   would be empty.

\noindent In this region, the agent density is not sufficiently high to  adopt the same cell-partition technique.
 The new approach exploits the  structure of the  paths performed by an agent that walks  for a long
 time in the Suburb and on the probability that she meets    agents coming from   the Central Zone.

 \noindent
 We say an agent performs a \emph{turn} when she changes direction during her Manhattan path.
 Let $\sa$ be an agent,  for any time $t$,     
  the random variable  $H_{t,\tau}$ counts  the number of turns performed by $\sa$ during the time interval $[ t, t + \tau]$.
  The next lemma shows that  this number  cannot be ``too'' large.


   \begin{lemma}\label{lem:suburb1}
  Let  $t \geq 0$ and let $\tau$ be such that $\frac{L}{n \sv} \leqslant \tau \leqslant \frac{L}{4 \sv} $.
   With probability at least $1- 1/n^4$, it
   holds that   
   
   \[ H_{t,\tau} \leqslant  \frac{4 \log n} { \log \left (  \frac L {\sv \tau}\right)} \]
 \end{lemma}
 
 \noindent
The previous lemma allows us to get high  probability for the existence of a ``good'' segment traveled by any agent
in the Suburb.

 \noindent  In the sequel,  we   assume that   agent $\sa$  lies in the South-West subsquare of size $L/2$, i.e.,  the subsquare
$[0,L/2] \times [0,L/2]$.  The $\sa$'s position 
  at time $t$ is  denoted as     $(x_0,y_0)$.
The analysis of the other three subsquares is symmetric.

\begin{lemma}\label{lem:suburb2}
Let  $t \geq 0$ and let $\tau$ be such that 
\[  \frac{ \max\{L/n, 4x_0,4 y_0 \}}{\sv}  \ \leqslant  \ \tau  \ \leqslant \   \frac{L}{4 \sv}\] Then, during the time interval
$[t,t+ \tau]$, 
 with probability at least $1 -1/n^4$, agent $\sa$  travels over a (horizontal or vertical)
segment directed to  the Central Zone and  that  has
length at least
\[  \frac{\sv \,  \tau \log \left (  \frac{L}{\sv \tau}\right)} {40 \log n  }    \]
\end{lemma}
 
\noindent
For the sake of convenience,    let  
$S  =  \frac {3  L^3 \log n} { 2\ell^2 n}$. The next lemma shows that the diameter of the SW Suburb is bounded by $S$.

\begin{lemma}\label{lem::sub00}
For every point $(x_0,y_0)$ in the south-west corner of the Suburb, it holds  that both $x_0$ and
$y_0$, are not larger than  $S$.
\end{lemma}

 \medskip
 \noindent \textbf{Meeting   agents coming  from the Central Zone.}
\noindent
Two   agents are said to \emph{meet} each other at time $t \geqslant 0$ if, at that time, their relative distance is not larger than
$(3/4) R$.  Observe that, due to Ineq. \ref{eq::SpeedBound},  if one informed agent meets another agent at some time, then within the next   
time unit,  the latter will get informed.

\noindent There may be some non-informed agents that travel over the Suburb    for a long period.
For those agents, the only chance to be informed (within relatively-small time) is to \emph{meet}
    agents coming from     the Central Zone. A symmetric argument will be applied to manage the case where the
    source will be in the Suburb for a long time.


\smallskip
\noindent
We say that a point  belongs to  the \emph{Extended Suburb} if the Manhattan distance 
between the point  and the Suburb is not larger than 
$2S$. Clearly, all points in the Suburb   belong to the Extended Suburb. 

\smallskip
\noindent
The  first property of the next Lemma 
is used when the source lies in Central Zone. 
 The second property is instead used  when  the source lies in the Suburb for a long  time.
   
\begin{lemma} \label{lm:meeting time}
   Let   $\sa$ be an agent   lying  in the Extended Suburb at any time $t \geqslant S / \sv$.
  For sufficiently large $n$, with probability at least $1-1/n^2$,  
   an agent $\sbb$ exists  that  has the following properties:
     \begin{enumerate}
\item  \label{lm:item1}
$\sbb$ was in the Central Zone at time $t- S / \sv$    and $\sbb$ 
   will meet  agent $\sa$
 within time $T + \tau$, where $\tau = 590 (S/\sv)$.
\item \label{lm:item2} Agent $\sbb$, after meeting $\sa$,  will be
  in    the Central Zone\footnote{It is not relevant
whether $\sbb$ will  visit the Suburb or not.} within time $t + \tau + 3 (S/\sv)$. 
\end{enumerate}

\end{lemma}

\proof
Let $(x_0,y_0)$ be the position of $\sa$ at time $t$. Observe that $\frac{L}{n}\leqslant S$ and,
 from Lemma \ref{lem::sub00} and the definition of Extended Suburb, both $x_0$ and $y_0$ are  not larger than $3S$.
Thus  it holds that  
\begin{equation}\label{terzo3}
 \frac{ \max\{L/n, 4x_0,4 y_0 \}}{\sv}  \ \leqslant  \frac{12S}{\sv} \ < \ \tau  
 \end{equation}
 From Ineq. \ref{primo1} and Ineq. \ref{eq::TransmBound} we obtain  
 \begin{equation}\label{utile}
 \frac{L}{\ell}\leqslant \frac{1+\sqrt{5}}{200}\sqrt{\frac{n}{\log n}}
 \end{equation}
 We thus  get
 \begin{equation}\label{secondo2}
\tau = 590 \frac{ S} {\sv}=\frac{L}{4\sv}\left(3540\frac{L^2\log n}{\ell^2 n}\right)\leqslant \frac{L}{4\sv}\left(\frac{3540\cdot (1+\sqrt{5})^2}{200^2}\right)
< \frac{L}{4\sv}\nonumber 
 \end{equation}
 Due to the above inequality  and Ineq. \ref{terzo3}, we can  apply Lemma \ref{lem:suburb2} with $\tau$ and $t$ specified in the  thesis.
 Hence, with probability at least $1 - 1/n^4$, agent $\sa$ will travel a \emph{good} segment (i.e. toward the Central Zone) of length
\[
d \ = \  \frac{ \sv \,  \tau \log \left (  \frac{L}{\sv \tau}\right)} {40 \log n  }
   \ \mbox{ during time interval }  \  [ T   \  , \ T +  \tau ]
\]
Observe that 
\begin{equation}\label{eq::goodseg}
d= \frac{590S\log\left(\frac{L}{590S}\right)}{40\log n}\geq 22\frac{L^3}{\ell^2n}\log\left(\frac{\ell^2n}{885L^2\log n}\right)
\end{equation}
\noindent
Wlog, we assume that the good segment is horizontal.
Let $t_\sa$ be the time   in which $\sa$ starts running the good segment and let $(x_\sa,y_\sa)$ be her position at that time. 
Consider the rectangle $I$ such that: its 
SW vertex is  point $(x_\sa + d + D, y_\sa)$ where $D = d/4 + \sv(t_\sa - t + S/ \sv)$, its horizontal size is $d/2$,
 and its vertical size is $\ell$.    The next claim (its proof is  in the Appendix) is the key-ingredient of the proof.
    
 
   {\small  \begin{quote}
\begin{claim} \label{cl::meet}
For sufficiently large $n$, with probability at least $1-1/n^2$, an agent  $\sbb$ exists   satisfying the following properties:

\begin{enumerate} 
\item \label{clm1}  at time  $t - S/\sv$
she is in some position  $(x_\sbb,y_\sbb) \in I$ and   has destination  $(x, y_\sbb)$,  for  some
  $ 0 \leqslant x \leqslant x_\sa +  d/2 $, and 
  \item  \label{clm2}  her  destination after next is in the Central Zone. \end{enumerate}
\end{claim}
 \end{quote}  }  

\noindent
We now show that the two properties of  Claim \ref{cl::meet} imply the two properties of the lemma.
 Observe that  Ineq.  \ref{eq::TransmBound} implies that 
    $I$ fully belongs to the Central Zone. Let $\sbb$ be an agent satisfying the two properties of Claim \ref{clm1} and
    let  
     $\bar t=\frac{x_\sbb-x_\sa+\sv t_\sa -S -\sv t}{2\sv}$; consider the horizontal coordinates $\bar x_\sa$ and 
     $\bar x_\sbb$,  at time $t + \bar t$,  of agents $\sa$ and $\sbb$, respectively. It holds that 
     
     \[ \bar x_\sa =   x_\sa +\sv(\bar t + t - t_\sa) =  \frac{x_\sa + x_\sbb - \sv(t_\sa - t + S / \sv)}{2} \]
     
      \[ \bar x_\sbb =   x_\sbb -\sv \left(t+ \bar t - \left(t-\frac{S}{\sv}\right)\right) =    \frac{x_\sa + x_\sbb - \sv(t_\sa - t + S /\sv)}{2} \]
     Moreover   observe that, by definition of rectangle $I$, it holds that

\[  x_\sa + \frac 58 d \leqslant \ \bar x_\sa \ = \ \bar x_\sbb  \ \leqslant x_\sa + \frac 78 d \] 
Hence   agents     $\sa$ and $\sbb$  at time $t+\bar t$
are at points 
$(\bar x_\sa,y_\sa)$ and $(\bar x_\sa , y_\sbb)$, respectively. 
   Their    distance at that time  is 
\[ |y_\sa-y_\sbb| \ \leqslant \  \ell \ \leqslant \ \frac{R}{\sqrt{5}} \leqslant \frac{3}{4}R \]
Hence, agents $\sa$ and $\sbb$ will meet  at time $t+\bar t$  and, moreover,  

\[   \bar t   =  \frac{x_\sbb-x_\sa+\sv t_\sa -S -\sv t}{2\sv}  \leqslant  \frac{(5/4) d + 2 \sv t_\sa - 2 \sv}{2\sv} =
 \frac{5 d}{8 \sv } + t_\sa -t  = \tau - \frac{3d}{8 \sv} < \tau \]
 This shows the first property of the Lemma.
 
 \noindent
 As for the second property, we observe that agent $\sbb$ will reach position $(x_\sa + \frac d2, y_\sbb)$ within
 time $t+\tau$. Two cases may arise.
 
 \begin{itemize}
\item Position   $(x_\sa + \frac d2, y_\sbb)$ lies in the Central Zone.\\
From the above observation,  we immediately have that 
  $\sbb$ will be in the Central Zone after meeting $\sa$ within
 time $t+\tau$.

\item Position   $(x_\sa + \frac d2, y_\sbb)$ lies in the Suburb. \\
Let $\hat x_\sbb$ be the horizontal coordinate of the $\sbb$'s destination determined in the first
property of Claim \ref{cl::meet}; it holds that
$ x_\sa + \frac d2 - \hat x_\sbb   \leqslant S - \hat x_\sbb  \leqslant  S$ \\
Hence, $\sbb$ reaches this destination within time $t + \tau +  S/\sv$.
The second property of Claim \ref{cl::meet} ensures that  the next destination is in the Central Zone.
It is easy to see that the maximal traveled distance to enter into the Central Zone is $2S$.
It follows that, within time $t + \tau +  3 (S/\sv)$, agent $\sbb$ will be in the Central Zone.

\end{itemize}

\qed\\

\medskip
\noindent {\bf Proof of Theorem \ref{thm:main}.}
Let time step $0$ be the starting time of flooding: we assume the MANET is already in its stationary phase.

\smallskip

\noindent
- We first consider the case where the source lies in the Central Zone when   flooding starts.  
From Theorem \ref{spreading}, with probability at least $1- 1/n^2$, at time $T_c = O(L/R)$,
all the cells of the Central Zone are informed. Observe 
that if \[ R \geq \frac{1 + \sqrt 5}{2}  L \left(\frac{3\log n }{ n}\right)^{1/3} \]
    Cor. \ref{cor::allcentr}  implies that the Suburb is empty and, hence, the flooding is completed.
     In the rest of the proof,  we can thus  assume that 
 Ineq.  \ref{eq::sub1} holds and focus only on those agents that at time $T_c$ are not in Central Zone.  
 Consider any agent $\sa$ among the latter agents.  By definition of Extended Suburb,  if an agent is not
 in the Central Zone at time $T_c$, then she will necessarily be in the Extended Suburb at time $T_c +
 S/\sv$.
  So,  by applying  Lemma \ref{lm:meeting time} to agent $\sa$ with $t = T_c + S/\sv$, we obtain that,
 with probability  at least $1 - 1/n^2$, an agent $\sbb$ exists that was in the Central Zone at time $T_c$
 and she will meet $\sa$ within time $T_c + O(S/\sv)$. Within the latter time, agent $\sa$ will be thus 
 informed with probability $1 - 1/n^2$.
 By using the union bound, we get that all such agents will be informed with high probability within time
 
 \[T_c + O\left( \frac{L}{\sv} \frac{L^2}{R^2} \frac{\log n}{n}\right) \ \  \ \mbox{
 since } \ \ S = \Theta (\frac{L^3 \log n}{ R^2 n}) \]

\noindent 
 -  We now consider the    case where      the source agent lies in Suburb 
 when   flooding starts.  By applying  Property 2 of Lemma \ref{lm:meeting time}
 to the source agent with $t = S /\sv$, we get that, with probability at least $1-1/n^2$,
 there is an agent $\sbb$ that meets the source agent and, after that, will be
  in the Central Zone within time $O(S/\sv)$.  The rest of the proof works as in the first case.

\qed

\section{A lower bound for flooding time }

We   observe that our upper bound holds for arbitrary small  agent speed $\sv$ while if 
 $\sv =0$,  flooding never terminates whenever the Suburb is not empty.
More generally, we   prove  (see the Appendix) the following lower bound

\begin{theorem} \label{thm::lb}
If $R = O(L/n^{1/3})$ then, with constant positive  probability,     
the flooding time is $\Omega(L/(\sv n^{1/3}))$. 
\end{theorem}

\noindent
Let us see when the above lower bound is 
  asymptotically larger than $L/R$. A necessary and sufficient condition is 
  $R/(\sv n^{1/3})  \rightarrow \infty$. 
From the theorem's hypothesis, this is true only if $L/(\sv n^{2/3})  \rightarrow \infty$. 
For instance, if $L = n^{1/2}$ then we  need $\sv$ asymptotically smaller  than $1/n^{1/6}$.

\noindent
If $L = n$  and  $R = L/n^{1/3} = n^{2/3}$ 
(so a large transmission radius) then  
the lower bound  above  becomes $\Omega(n^{2/3}/\sv)$:  for $\sv  = \Theta(1)$, this is  
larger than $L/R$ for an  $n^{1/3}$ factor.

\bigskip
\noindent
{\large \textbf{Acknolewdgements.}} We thank Francesco Pasquale for helpful comments on a preliminary
version of this paper.

\newpage


\newpage

 \appendix
 
\section{Proofs for the Central Zone}

\subsection{Proof of Lemma \ref{boundary}}
 
For the sake of convenience, we say that a cell in $B$ is a black cell and all
 the cells not in $B$ are white cells. In the sequel, with the term  \emph{row} (\emph{column}) we   mean  the subrow (subcolumn)  of cells
that belong  to the Central Zone. Observe that, according to this definition, rows and columns have variable length.
We say that a row  is black if all the cells of the row are black. Similarly, we define
a black column. Moreover, a row or a column which contains both at least one black cell and at least one
white cell is said to be gray.

\noindent
We   easily observe that the following inequalities hold

\begin{equation} \label{Eq::cz1}
\sqrt{\min\{|B|, |\sCZ| - |B|\}}  \ \le \ \sqrt{\frac { |\sCZ| }{2} } \ \le \ \frac { m}{\sqrt 2}
\end{equation}

\noindent Let $b_r$ and $b_c$ be, respectively, the number of black rows and the
 number of black columns.
In order  to prove the lemma,  we distinguish four cases.
\begin{description}
\item[$b_r = 0 \wedge b_c \geqslant 1$:] In this case, from Lemma \ref{obs::38}, the number  of gray rows
is at least $m / \sqrt 2$. This implies   every gray row contains at least one cell in
$\partial B$. Then, $ |\partial B | \geq m / \sqrt 2$  and the lemma follows from Ineq. \ref{Eq::cz1}.

\item[$b_r \geqslant 1 \wedge b_c = 0$:] This case is symmetric to the previous one.

\item[$b_r \geqslant 1 \wedge b_c \geqslant 1$:]  If there exist a black row  and a black  column of maximal length $m$,  then all non-black rows and 
columns  are gray. Observe that each of the $|\sCZ|-|B|$ white cells uniquely determines a pair formed by one gray row and one gray column.  We thus  get

\[    |\sCZ|-|B|      \le   (m -b_c)(m-b_r)  \]
Wlog, we assume that $b_r \le b_c$, then the above inequality implies that

\[    |\sCZ|-|B|  \le (m-b_r) \]  The thesis follows from the fact that $|\partial B| \ge m - b_r$.
Now,  we consider the case where no black row or black column of maximal length do exist. Wlog, we assume that no black row of maximal length exists. Since there are at least a black column, then all rows of maximal length are gray. From Lemma  \ref{obs::38}, the number  of gray rows
is at least $m / \sqrt 2$. This implies   every gray row contains at least one cell in
$\partial B$. Then, $ |\partial B | \geq m / \sqrt 2$  and the lemma follows from Ineq. \ref{Eq::cz1}.

\item[$b_r = 0 \wedge b_c = 0$:] Let $y_r$ and $y_c$ be, respectively, the number of gray rows and the number of gray columns.
Since there are neither black rows nor black columns, it must be the case that every black cell belongs to both a gray row and
a gray column. This implies that
\[ y_r\cdot y_c \;\geqslant |B|. \]
Without loss of generality, assume that $y_r \geqslant y_c$. It follows that $y_r^2 \geqslant |B|$, and so
$y_r \geqslant \sqrt{|B|}$. Since every gray row contains at least a cell in $\partial B$, it holds that $|\partial B|
\geqslant \sqrt{|B|} \geqslant \sqrt{\min\{|B|, m^2 - |B|\}}$.
\end{description}
\qed

\section{Proofs for the Suburb}

\subsection{Proof of Lemma \ref{lem:suburb1}}


 \noindent
For the sake of simplicity,  we will use the following probability notations.
 For an event $\mathcal{E}$ and a r.v.
$X$,  the notation $ \Probc{\mathcal{E}}{X} \ \leqslant   p$
means that, for every possible value $x$ of $X$, it holds
$
\Probc{\mathcal{E}}{X=x}  \leqslant      p
$.
 
 \noindent Consider the turns performed by $\sa$ after time $t$.
 For any $i = 1, 2, \ldots $, define  $X_i$ the distance travelled by agent $\sa$ between the $i$-th turn
 and the $i+1$-th turn.   We then consider the binary  r.v.  defined as follows

 \[  Y_i  =   1 \  \mbox{ if  }   \ X_i \  \leqslant   \ \sv \tau \ \mbox{ and } 0  \ \mbox{ otherwise } \]

 Observe that if $X_i \leqslant d$ then the $(i+1)$-turn point lies in the square centered  at the $i$-th turn point, 
 with diagonals parallel to the axis, and that has side length  $\sqrt 2 \, \sv \tau$.
 So, it holds that

 \[   \Probc{Y_i =1}{X_1, \ldots , X_{i-1}}         \  \leqslant \ p \]
 
 \[ \mbox{ where } \ p \ =  \
  \frac{ (\sqrt 2 \, \sv \tau)^2 } {L^2} \]
  Notice that, since $\tau \leqslant L / (4 \sv)$, then $p < 1$.
  We now need the following  standard probability bound (See~\cite{ABKU99}).

\begin{small}
\begin{quote}
\begin{claim}\label{lemma:ba}
Let $X_1, \dots, X_n$ be a sequence of
 random variables with values in an arbitrary domain, and let $Y_1, \dots, Y_n$ be
 a sequence of binary random variables, with the property that $Y_i = Y_i(X_1, \dots, X_i)$. If
\[
\Prob{Y_i = 1 \;|\; X_1, \dots, X_{i-1}} \leqslant p
\]
then
\[
\Prob{\sum Y_i \geqslant k} \leqslant \Prob{B(n,p) \geqslant k}\] where $B(n,p)$ denotes the binomially distributed random variable with parameters $n$ and $p$.
\end{claim}
\end{quote}
\end{small}

\noindent
For any $ h = 1,2, \ldots$,
  from  the above Claim it holds that 
  
   \[ \Prob{\sum_{i =1}^h  Y_i  \ =   \  h   }  \ \leqslant \ \Prob{B\left(h, p \right) = h   } \]
   So, it clearly holds that
   
     \begin{equation} \label{eq::B}
      \Prob{\sum_{i =1}^h  Y_i  \ =   \  h   }  \ \leqslant \ p^h
    \end{equation}

      \noindent
We observe that
\[   \sum_{i = 1}^{H_{t,\tau} -1}  X_i  \ \leqslant  \sv \tau \]  Hence,  event `` $H_{t,\tau}  \geqslant h+1$ ''
implies event ``$ \sum_{i = 1}^{h} X_i \leqslant \sv \tau$'' and so

\[   \Prob{H_{t,\tau}  \geqslant h+1  }  \  \leqslant  \  \Prob{   \sum_{i = 1}^{h} X_i \leqslant  \sv \tau  }
\]
Moreover,  observe that the following implication holds
\[   \sum_{i = 1}^{h} Y_i  <    h  \  \Longrightarrow  \  \sum_{i = 1}^{h} X_i  \  > \  \sv \tau \]
It thus follows that  also the implication below holds 
  \[ \sum_{i = 1}^{h} X_i \ \leqslant  \sv \tau   \  \Longrightarrow  \  \sum_{i = 1}^{h} Y_i  \  =  \    h   \]
Hence, from Ineq.  \ref{eq::B},   we get

\begin{equation} \label{eq::HH}      \Prob{H_{t,\tau}  \geqslant h+1  }    \ \leqslant  \  \Prob{   \sum_{i = 1}^{h} X_i \leqslant  \sv \tau }    \leqslant
\Prob{\sum_{i =1}^h  Y_i =  h   }      \leqslant  \ p^h  \end{equation}
Observe that for $h = \left \lceil   \frac{4 \log n}{\log (1/p) }\right \rceil$ it holds that

\[       \Prob{H_{t,\tau}  \geqslant h+1  }  \ \leqslant \ \frac 1 {n^4}           \]
We also have that

\[  \frac{4 \log n}{\log (1/p)} \ =  \ \frac{4 \log n}{ \log\left (\frac{L^2}{(\sqrt 2 \sv \tau)^2} \right)} \ =
\ \frac{2 \log n}{ \log\left (\frac{L }{\sqrt 2 \sv \tau } \right)} \]
Hence, with probability at least  $1 - 1/n^4$, it holds that
\begin{eqnarray*}
H_{t,\tau} & \leqslant & \left\lceil \frac{2 \log n}{ \log\left (\frac{L }{\sqrt 2 \sv \tau } \right)}\right \rceil  \ \leqslant \
1 + \frac{2 \log n}{ \log\left (\frac{L }{\sqrt 2 \sv \tau } \right)} \\
&  =  &  \left ( 1 + \frac{2 \log n}{ \log\left (\frac{L }{\sqrt 2 \sv \tau } \right)} \right ) \frac{4 \log n }
{\log\left (\frac{L }{ \sv \tau } \right) }  \frac { \log\left (\frac{L }{ \sv \tau } \right) }{4 \log n }  \\
& = &  \frac{4 \log n }
{\log\left (\frac{L }{ \sv \tau } \right) }  \left ( \frac { \log\left (\frac{L }{ \sv \tau } \right) }{4 \log n } + 
\frac{ 2  \log\left (\frac{L }{ \sv \tau } \right)}{ 4 \log\left ( \frac{L }{ \sqrt 2 \sv \tau } \right) } \right) \\
 \mbox{ (since } \   \frac L {\sv \tau} \leqslant  n) & \leqslant &  
 \frac{4 \log n }
{\log\left (\frac{L }{ \sv \tau } \right) } \left( \frac  {1 }{4   } + \frac 24 
\frac{    \log\left (\frac{L }{ \sv \tau } \right)}{   \log\left ( \frac{L }{   \sv \tau } \right) - \frac 12} \right)
\end{eqnarray*}

\noindent
Notice that

\[ \frac{    \log\left (\frac{L }{ \sv \tau } \right)}{   \log\left ( \frac{L }{   \sv \tau } \right) - \frac 12} 
\ = \ \frac 1 {1 - \frac{1}{2 \log\left (\frac{L }{ \sv \tau } \right)}}  \ \leqslant \ \frac 4 3 \ \
\mbox{ (since } \   \frac L {\sv \tau} \geqslant  4) \]
\noindent Finally, we get

\[ H_{t,\tau}   \ \leqslant \    \frac{4 \log n }
{\log\left (\frac{L }{ \sv \tau } \right) } \left( \frac  {1 }{4   } + \frac 24 \cdot
\frac 43 \right) \ \leqslant \   \frac{4 \log n }
{\log\left (\frac{L }{ \sv \tau } \right) }\]

 \qed

\subsection{Proof of Lemma  \ref{lem:suburb2}}

 Let $k= H_{t,\tau}$   be
 the number of  \sa's  turns in the interval   $[t,t+\tau]$. For any $i = 1, \ldots , k$, we define
 $(x_i,y_i)$ as the $i$-th turn position of agent \sa.  We denote the \sa's position at time $t+ \tau$
 as $(x_{k+1},y_{k+1})$. For any $j = 1, \ldots , k+1$, define $h_j = x_j - x_{j-1}$ and $v_j =  y_j - y_{j-1}$.

\noindent
Observe that when $h_j$ or $v_j$ are positive then  the travelled segment is directed towards the Central Zone.
It holds that

\[    \sum_{j=1}^{k+1}  | h_j|  +     \sum_{j=1}^{k+1}  | v_j|   =  \sv \,  \tau \]
Wlog, we assume that the first sum is not smaller than the second one. So,

\[    \sum_{j=1}^{k+1}  | h_j|  \geqslant \frac{\sv \tau }{2}  \]
Now, define the following index subsets:

\[  J^+ =  \{  j \ | \ h_j > 0 \} \  \mbox{ and } J^- =  \{  j \ | \ h_j \leqslant  0 \} \ \]
Hence, we get

\begin{equation} \label{eq::HJ}
\sum_{j \in J^+} h_j  \geqslant  \frac{\sv \tau }2  + \sum_{j \in J^-} h_j
\end{equation}

\noindent
Observe that

\[  \sum_{j \in J^-} h_j +  \sum_{j \in J^+} h_j  =  \sum_{j=1}^{k+1}   h_j = x_{k+1}  - x_0 \geqslant
- x_0 \, \geqslant \, -  \frac {\sv \, \tau} 4  \]  
This implies

\[  \sum_{j \in J^-} h_j  \geqslant - \frac {\sv \, \tau} 4 -    \sum_{j \in J^+} h_j   \]
By combining the above equation with Eq. \ref{eq::HJ}, we obtain

\[    \sum_{j \in J^+} h_j  \geqslant  \frac{\sv \tau   }2   - \frac{\sv \tau }4     - \sum_{j \in J^+} h_j
 \] So, 
 \[   \sum_{j \in J^+} h_j   \geqslant   \frac{\sv \tau   }8 \]
 Hence, we can say that an index $\hat j$ exists  such that

 \[   h_{\hat j }   \geqslant  \frac{ \sv \tau   }{  8( k +  1) } \] From Lemma  \ref{lem:suburb1}
 and the fact that $\tau \geqslant L /(n \sv)$,  with probability at least $1 - 1/n^4$  it holds  that
 
  \[      h_{\hat j } \    \geqslant    \frac{\sv \tau}{ 8\left(  \frac {4 \log n }{ \log\left( \frac L {\sv\tau}\right)   } +1 \right) }   
  \geqslant \   \frac{\sv \,  \tau \log \left (  \frac{L}{\sv \tau}\right)} {40 \log n  } \]
 
\qed

\subsection{Proof of Lemma \ref{lem::sub00}}
 Let $(\overline x,\overline y)$ the SW corner of the cell containing $(x_0,y_0)$.
We will prove the lemma's bound   for $x_0$;  the bound for $y_0$ can be obtained by a symmetric argument.
 Since this  cell does not belong to the Central Zone, by Obs. \ref{obs::cellprob}, we get
 
\[      \frac{3\ell^2}{L^4} \left(   \frac {\ell} 3  (3L - 2\ell) 
+ \overline x (L-\ell - \overline x) +  \overline y (L -\ell -\overline y) \right)  
\ \leqslant \ \frac{3}{8}  \frac{\log n}n \]
From the above inequality, we get

\begin{equation} \label{eq::b1}
\overline x (L - \ell - \overline x)  \ \leqslant \ \frac{ L^4 \log n}{8 \ell^2 n}
\end{equation}

\noindent
Notice that, by definition of $\ell$ and  Ineq. \ref{eq::sub1},  we get

\begin{equation}\label{eq::b2}
\ell \ \leqslant \ \frac{R}{\sqrt 5} \ \leqslant \ \frac 3 {40} \, L
\end{equation}

\noindent
From Lemma \ref{obs::38}, we obtain 

\[ \overline x \ \leqslant \ \frac{m-\frac m {\sqrt 2}} {2 } \, \ell = \frac{2 - \sqrt 2} 4 \, L\]
From this inequality and Ineq. \ref{eq::b2}, we obtain

\[   \L - \ell - \overline x \ \geqslant  \  L -  \frac 3 {40} \, L - \frac{2 - \sqrt 2} 4 \, L \ \geqslant \ \frac 3 4 L \]
From the above inequality and Ineq. \ref{eq::b1}, it holds that
\[ \frac 3 4 \, \overline x L \leqslant \overline x (L - \ell - \overline x)  \ \leqslant \ \frac{ L^4 \log n}{8 \ell^2 n}     \]
and
\begin{equation} \label{eq::b3}
\overline x \ \leqslant \  \frac{  L^3 \log n}{6 \ell^2 n}  
\end{equation}

\noindent
Since  $x_0 \leqslant \overline x + \ell$, we now bound $\ell$.  
From Ineq. \ref{eq::sub1},  we obtain

 \[ \ell    \leqslant   \frac R { \sqrt 5} \ \leqslant   
  \frac{1 + \sqrt 5}{2\sqrt 5}  L \left(\frac{3\log n}{n}\right)^{1/3}   \]
 Then   
   \[   \ell^3 \    \leqslant    \
  \left(\frac{1 + \sqrt 5}{2\sqrt 5} \right)^3  3  L^3  \frac{\log n}{n}  \]
  We thus  get 
   
   \[   \ell \ \leqslant \   \left( \frac{1 + \sqrt 5}{2\sqrt 5} \right)^3 \frac{  3 L^3 \log n}{  \ell^2 n}
   \ \leqslant \ 1.2 \, \frac{    L^3 \log n}{  \ell^2 n} \]
   From Ineq. \ref{eq::b3} and the last one, we finally obtain
   
   \[   x_0 \ \leqslant \ \overline x + \ell \ \leqslant \  \frac{1}{6} \frac{  L^3 \log n}{ \ell^2 n}  + 
   1.2 \, \frac{   L^3 \log n}{  \ell^2 n} \ \leqslant \ \frac{ 3 L^3 \log n}{2 \ell^2 n}=S \]
 \qed


\subsection{Proof of Claim \ref{cl::meet}}

    
   Let $\bar P(x_\sbb,y_\sbb)$ be the probability that an agent $\sbb$, being in $(x_\sbb,y_\sbb)$, 
has destination  $(x, y_\sbb)$  for  some
  $ 0 \leqslant x \leqslant x_\sa +  d/2 $.
 From Eq. \ref{eq::fiwest}, it holds that

\[\bar P(x_\sbb,y_\sbb) = \frac{x_\sa +  d/2}{x_\sbb}  \phi^{\mbox{W}}_{(x_\sbb,y_\sbb)}  \ = \ 
\frac{(x_\sa +  d/2)  (L-x_\sbb) }{ 4L(x_\sbb+y_\sbb) - 4(x_\sbb^2 + y_\sbb^2)} \]

\noindent
Let $P_\sbb$ be the probability that agent $\sbb$ satisfies Property \ref{clm1}. Then,

\[ P_\sbb  = \int_I \bar P(x,y) f(x,y) \dx \dy \] where  $f(x,y)$ is  the probability density function of the  spatial distribution in Eq.
\ref{rpd}.   It thus follows that

\begin{eqnarray}\label{cinque5}
P_\sbb &  = & \int_I \frac{(x_\sa +  d/2)  (L-x) }{ 4L(x+y) - 4(x^2 + y^2)} \left( \frac{3}{L^3}(x+y)-\frac{3}{L^4}({x}^2+{y}^2) \right) \dx \dy \nonumber\\
& = &  \frac{3(x_\sa + d/2)}{4 L^4 } \int_I (L-x) \dx \dy \nonumber \\
& = &   \frac{3d \ell (x_\sa + d/2) (L - x_\sa - (5/4)d - D)}{8 L^4 } \nonumber\\
& \geq &   \frac{3d^2 \ell}{16 L^4 }  (L - x_\sa - (5/4)d - D)
\end{eqnarray}

\noindent
Observe that

\[\begin{array}{lcll}
x_\sa + (5/4)d + D  &=& x_\sa+ (3/2) d  + \sv (t_\sa - t + S/\sv)\\
&\leqslant &  x_\sa+ (3/2) d +( \sv \tau - d + S)& \mbox{Since $x_\sa \leqslant x_0 + \sv \tau - d $ }\\
&\leqslant & x_0+ 2\sv \tau +S &\mbox{By Lemma \ref{lem::sub00} }   \\
&\leqslant & 3S+ 2\sv \tau \\
&\leqslant& \frac{1775 L^3\log n}{\ell^2 n}&\mbox{By Ineq. \ref{utile} } \\
&\leqslant &\frac{L}{2} 
\end{array}
\]

\noindent
From the above inequality, Ineq. \ref{cinque5} and \ref{eq::goodseg}, we obtain  
\begin{equation}
P_\sbb \geq  \frac{3d^2 \ell}{32 L^3 }\geq 45\frac{L^3}{\ell^3n^2}\log^2\left(\frac{\ell^2n}{885L^2\log n}\right)
\end{equation}  

\noindent
It is easy to verify that the right-hand side of the above inequality is a decreasing function of $\ell$.  
  So, in order to get a lower bound for that value, we evaluate it in an upper bound of $\ell$.
  
  Now from Ineq.s \ref{primo1} and \ref{eq::sub1} we have that 
  \[ \ell \leqslant  \frac{R}{\sqrt{5} }
  \leqslant \frac{(1 + \sqrt 5)}{2\sqrt 5}  L \left(\frac{3\log n}{ n}\right)^{1/3}\leqslant \frac{16L}{15}\left(\frac{\log n}{n}\right)^{1/3} = \ell_{ub} \]
   Thus we have 
\begin{eqnarray}\label{otto8}
P_\sbb &\geq &  45\frac{L^3}{\ell_{ub}^3n^2}\log^2\left(\frac{\ell_{ub}^2n}{885L^2\log n}\right)\nonumber\\
&\geq & \frac{37}{n\log n}\log^2\left(\frac{13}{10^4}\left(\frac{n}{\log n}\right)^{1/3}\right)\nonumber\\
&\geq & \frac{37}{n\log n}\log^2\left(n^{1/4}\right)\geq  2.3\frac{\log n}{n} \,\,\,\,\,\,\,\,\, \mbox{For sufficiently large $n$.}\nonumber
\end{eqnarray} 

 \noindent
Since every  destination is selected uniformly at random over the 
 square and the Central Zone's area is 
(by Ineq.  \ref{eq::TransmBound}) at least $(11/12) L^2$, the probability that   agent $\sbb$ satisfies both properties of the claim 
is 
\begin{equation} \label{eq::probb}
P \geqslant \frac{11}{12} P_\sbb  \geqslant 2.1 \frac{\log n}{n}\nonumber 
\end{equation}

\noindent
Since there are $n-1$ independent agents, the probability that   no agent   satisfies both properties is 

\begin{eqnarray}
   (1- P)^{n-1}\leqslant  \left(1-2.1\frac{\log n}{n}\right)^{n-1}=e^{-2.1\log n \frac{n-1}{n}}\leqslant \frac 1 {n^{2}} \nonumber 
\end{eqnarray} 
where the last inequality holds for sufficiently large $n$.  

\qed

\section{The Lower Bound}

\textbf{Sketch of the Proof of Theorem \ref{thm::lb}.}
Let $d$ be such that $d = \Theta(L/n^{1/3})$ and $d \geqslant R$ (since 
$R = O(L/n^{1/3})$, such a $d$ does  exist). Let $F$ and $E$ be the subsquares having 
their SW corner in $(0,0)$ and side length $d$ and $3d$, respectively. 
By Observation \ref{obs::cellprob}, it holds that: 
the probability that a fixed agent lies in $F$ is 
$P_F = \Theta((d/L)^3)$ 
and the probability that a fixed agent lies in $E$ is 
$P_E = \Theta((d/L)^3)$. 
Consider the event $B =$ ``at time 0, at least an agent is in $F$ and 
no agent is in $E - F$''. 
Let $P$ be the probability that event $B$ holds. 
Then,
 \[ P \geqslant  \sum_{i = 1}^n \Prob{\mbox{agent } \ i \ \mbox{ is in } F } \Prob{\mbox {all  agents are 
not in }  E - F} 
= n P_F (1 - P_E)^{n - 1} 
= \Theta(1) \]
Hence, $P$ is a constant positive probability. 

\noindent
If event $B$ holds (and the 
source is not in $F$), an agent $\sa$ in $F$, at time 0, gets informed 
at time $t$ only if there is an informed agent   that, 
at time $t$, is at distance at most $R$ from $\sa$. Since at time 0 the 
distance from $\sa$ and any agent not in $F$  is at least $2d$, it takes 
at least a time span of $(2d - R)/(2\sv)$  so that   $\sa$  and an agent 
that was outside $E$ could be at distance not larger than $R$. Thus, 
the flooding time is at least 
$(2d - R)/(2\sv) = \Omega(L/(\sv n^{1/3}))$. 
\qed

\end{document}